\documentclass[a4paper]{jpconf}
\usepackage{graphicx}

\newcommand{\pT}{$p_T$}
\newcommand{\upctopo}{UPC\_Topo~}
\newcommand{\upcmain}{UPC\_Main~}

\newcommand{\snn}{$\sqrt{s_{_{\rm NN}}}$}

\begin{document}
\title{Rho Meson Diffraction off Au Nuclei}

\author{R. Debbe for the STAR Collaboration}

\address{Brookhaven National Laboratory Upton NY 11973, USA}

\ead{debbe@bnl.gov}

\begin{abstract}
The STAR Ultra Peripheral Collisions program has collected a substantial sample of $\rho$ mesons and for the first time at RHIC energies
it has been able to extract the distribution of momentum transfer t from diffractive elastic scattering off the Au ion. The resulting
diffraction pattern is consistent with coherent scattering off a nuclear object the size of the Au nuclei. Measurements of this nature 
can offer insights and guidance to the ongoing preparations for the new electron ion programs.
\end{abstract}

\section{Introduction}

Our perception of nuclei has evolved drastically since the early days of Rutherford's experiments which made them appears as almost point like
and massive objects. Higher energy electron scattering refined that knowledge providing the spatial distribution of charge in nuclei.
The actual extent of the strong force was investigated with beams of pions and protons but that program suffered from the difficulties 
in disentangling
the interference between Coulomb and nuclear interactions. At that point nuclei were well described by an almost spherical distribution
refered to as the Wood-Saxon distribution. With the access to much higher energies, electron scattering opened a whole new way to study
nucleons as it made accessible the study of their composite partonic nature. In particular, the HERA program identified the surprising
fact that up to 20\% of the interactions had the target nucleon left intact. This fact can be understod as the result of these high
energy interactions being able to resolve partons which carry even smaller fractions of the nucleon momentum and are subject to 
non-linear effects.
The ideal probe to study the nuclear component of  nuclei would be one that does not carry charge. Such probe can be found in the photon
fluctuations; a photon of enough energy can be itself or it may fluctuate into neutral vector mesons, quark pairs or lepton pairs with 
corresponding decreasing probability. Such beams of photons and its fluctuations are available at the RHIC and LHC colliders. High 
fluxes of photons described with the Weizs\"{a}ker-Williams formalism 
\cite{Bertulani}
 are present at RHIC  with energies that range up to 24 GeV in the photon-nucleon center of mass. 

We report the measurement of the transverse momentum of the $\rho^{0}$ meson after it undergoes an elastic scattering on the
corresponding Au ion. For the first time, a diffraction
pattern with several secondary peaks is clearly visible. Previous similar measurements were performed at low energy at SLAC \cite{busza} 
and at DESY with
bremsstrahlung photons \cite{samTing}, in both cases the distributions of transverse momentum of the $\rho$ mesons only shows the
main diffraction peak.     
This measurements come as preparations are taking place for a new Electron Ion Collider (EIC) that will revolutionize, one more time,
our ability to study nucleons and nuclei. This new program will, in addition to existing variables used to characterize the state of partons,
 include information about their distribution in the transverse plane. That information will be extracted from distributions similar
to the ones reported in this work as the transverse momentum of the recoil target is a conjugate of the impact parameter of the collision.

\begin{figure}[h]
\begin{minipage}{16pc}
\includegraphics[]{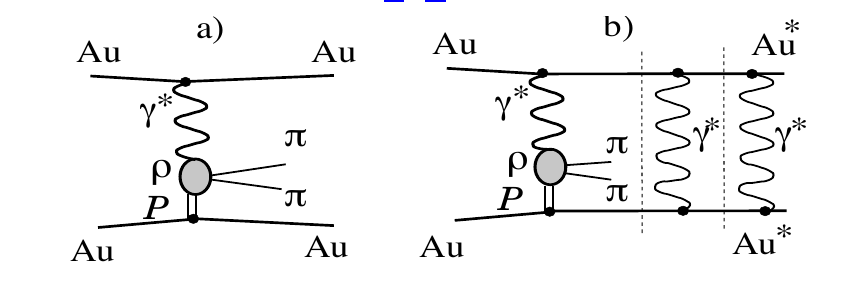}

\caption{\label{UPCdiagram}Schematic diagrams for UPC events. Panel a shows the most unbiased events. Panel b
shows UPC events  with further photon exchanges that exite both nuclei.}
\end{minipage}\hspace{4pc}%
\begin{minipage}{17pc}
\includegraphics[width=16pc]{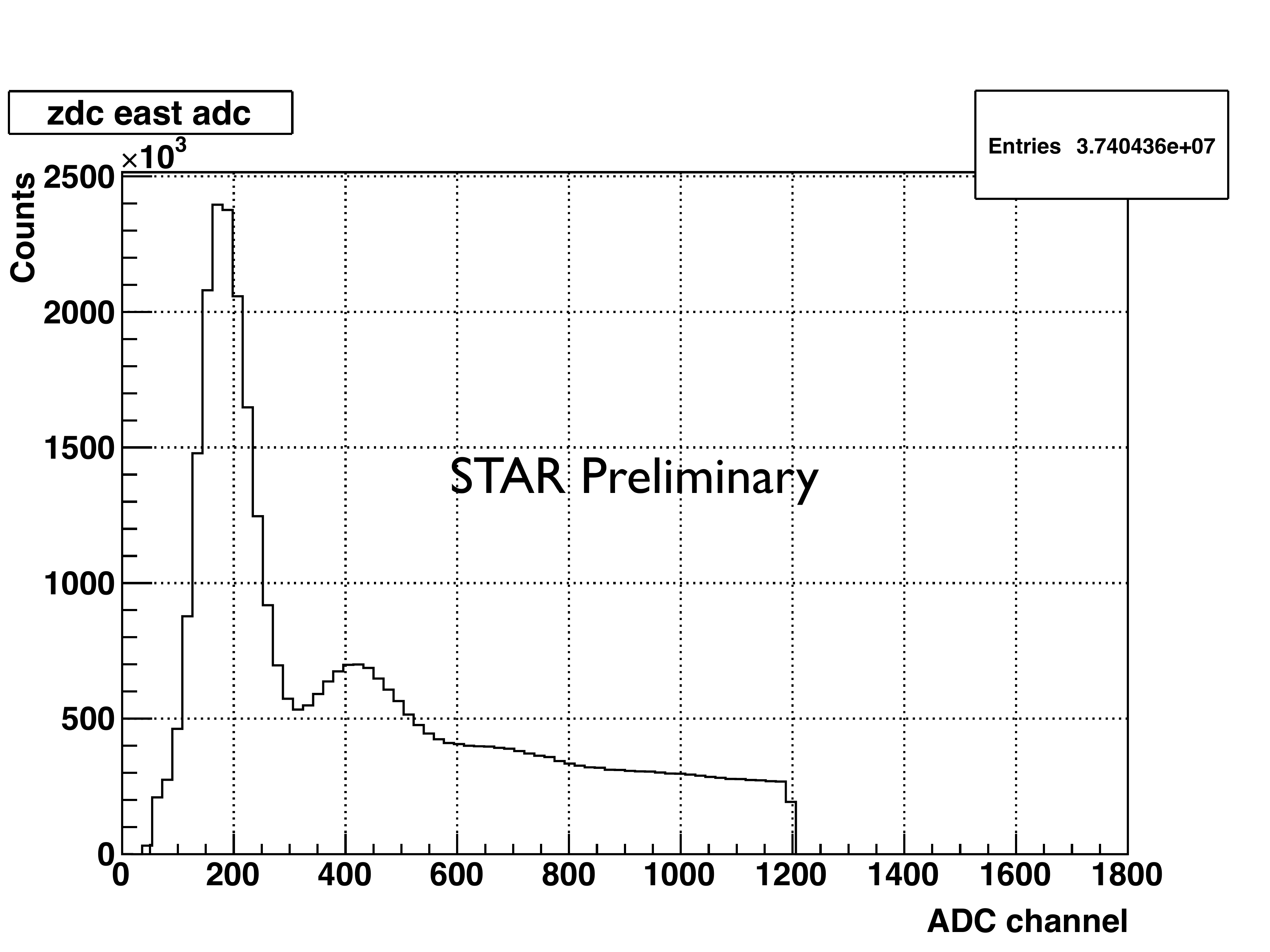}
\caption{\label{zdcCuts}The distribution of charge collected in one of the ZDC detectors, a clear one neutron 
peak is visible followed by a smaller one related to two neutrons.}
\end{minipage} 
\end{figure}

\section{The STAR Experimental setup}

The data used for this analysis were collected with the STAR detector at RHIC from Au+Au  collisions with center of mass energy per nucleon 
 \snn=200 GeV.
The momentum of charged particles were detected with the STAR Time Projection Chamber (TPC) in two units of pseudo-rapidity centered around 0
($| \eta| <1$) and full azimuthal coverage. The TPC records up to 45 samples of the ionization left in the detector gas which allows for
a good resolution particle identification based on energy loss. The charge left by particles inside the TPC drifts along its axis
and is read out on the East and West sides which are divided into six sectors each. The TPC is also used to identify the vertex of the 
collisions and together
with the bending power of the 0.5 Tesla magnetic field, it provides a momentum resolution equal to $\Delta p_{T}/p_{T} = 0.005 + 0.004 p_{T}$.
More details about the TPC can be found in \cite{TPCnim}.
The cylindrical TPC is completely surrounded by the Time Of Flight (TOF) detector consisting of 23040 Resistive Parallel Plate gas detectors
arranged in cells, groups of which form  modules installed in two sets of trays along the East and West sides of the TPC, 10 TOF trays
overlap the azimuth coverage of one TPC readout sector. The TOF detector was used to trigger the UPC events and provides good 
time-of-flight measurements, although this analysis doesn't make use of that information.
STAR has two Zero Degrees Calorimeters (ZDC) installed at $\pm 18$ meters away from the nominal interaction point. These calorimeters
are optimized for the detection of beam energy neutrons. These detector are instrumental in the definition of the trigger used for this analysis,
more details about them can be found in \cite{Adler:2001fq}.

\subsection{ Triggers}

At the time the data used for the present analysis was collected, two triggers were defined to collect events described by the
diagrams of Fig. \ref{UPCdiagram}. The event shown in panel a of that figure are collected with a trigger named \upctopo and events
similar to those depicted in panel b were collected with the so called \upcmain trigger. For both triggers the TOF detector is used to 
set limits 
in the number of tracks detected in
the event. The ZDC calorimeters are used to identify events where the nuclei undergo a Coulomb exitation (a Giant Dipole Resonance GDR) 
decaying 
into one or more 
neutrons emmitted along the beam. The Beam Beam Counter (BBC) is used as a veto of events with any activity at high rapidity.  
The \upctopo trigger is the most inclusive of the triggers and is designed to tag
the least biased UPC events; the ones without constraints on the number of neutrons detected in the ZDC calorimeters. This trigger
is based on the TOF detector ability to find a pair of back-to-back particles by selecting events with TOF hits in two azimuthally
opposite sectors (as mentioned above, a sector is formed with 10 trays). To avoid contamination from cosmic rays impinging 
vertically on the TPC the top (0) and bottom (3) sector are used as vetoes. 

The cross section for $\rho$  photo\-production tagged by \upctopo trigger  is $\sim 10$ times higher than the one for the \upcmain events 
\cite{baltzStarLight}, but the less restrictive nature of the first trigger makes it inefficient as it collect many spurious events, 
in contrast the \upcmain trigger is cleaner and more efficienct because of the conditions imposed on the number of neutrons 
detected in the ZDC detector. 

\subsection{Data sets}

This report is based on the analysis of 22.9 million \upcmain events written to tape by the STAR data aquisition system. This represents 
two thirds of the total number of events collected that year. (Events from an earlier part of that run period may be included in 
 future analyses once further tests are conducted  on the appropriate setup of the trigger.) Once the data is fully reconstructed a 
first selection of events is made. During run 10
the luminosity of RHIC was high enough such that the TPC being a slow detector had recorded several collisions (it takes $\sim 36~ \mu$ 
sec for the ionization left by a track close to the middle of the detector to drift to the readout planes). The events that are read in 
this first pass can contain
several vertices. The analysis of each events starts by looping over the vertices, pairs of 
 tracks originating from the same vertex are formed whenever the vertex was constructed out of at most 12 tracks. These pairs are then 
written to root tree files. Each track in the pair carries all the information extracted from the TPC
as well as the corresponding TOF relevant variables whenever available. Information about energy deposited in the EM calorimeter cells
is also written for each track. 
The root tree files are then processed separately to identify the events that where identified $\rho$ mesons were produced exclusively
in events that recorded a single neutron in the ZDC calorimeters. 

\section{The $ \rho^{0}$ meson identification} 

\begin{figure}[h]
\begin{minipage}{16pc}
\includegraphics[width=18pc]{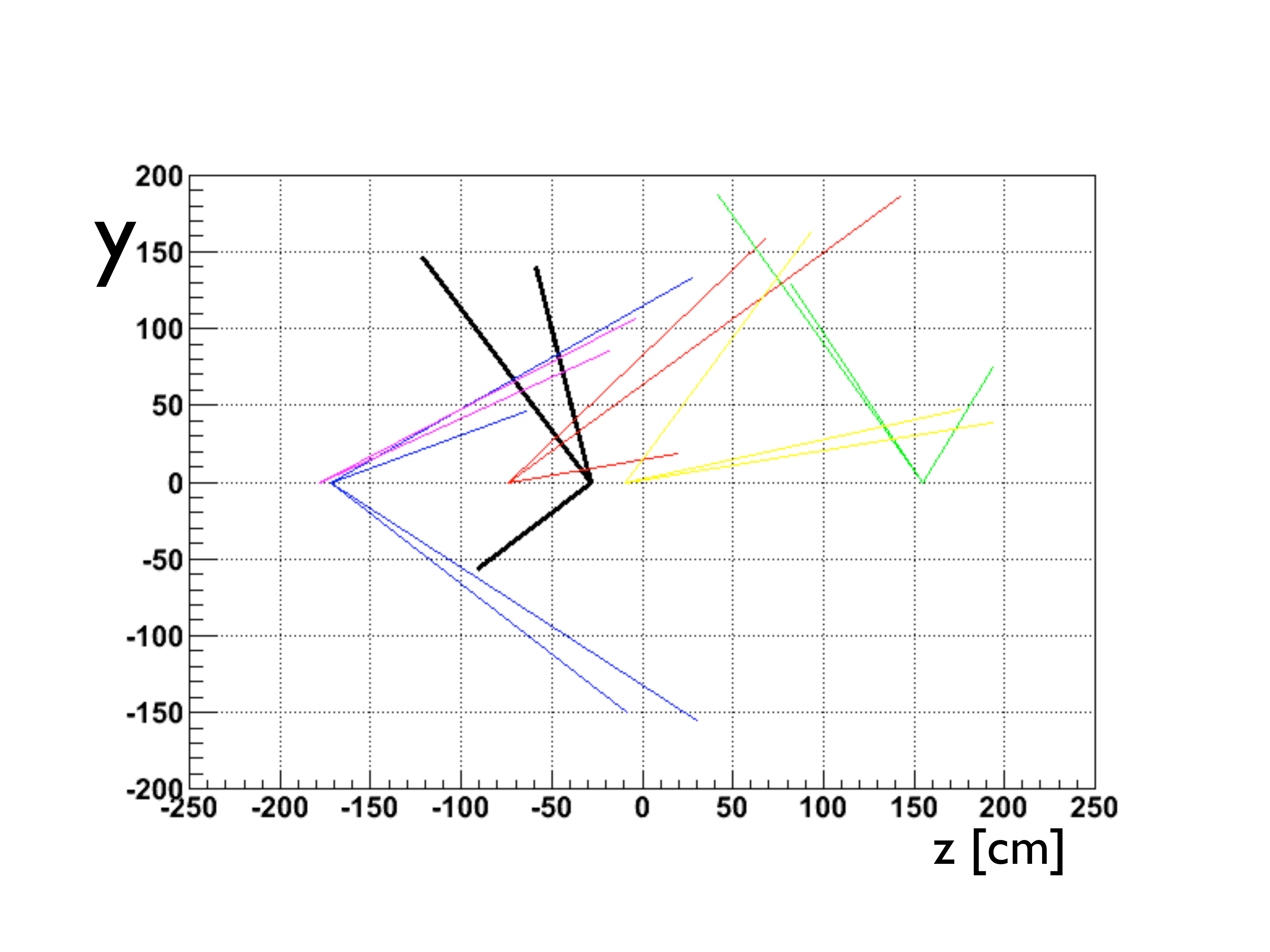}
\caption{\label{manyEvents}The projection of tracks on the YZ view of the STAR TPC is shown with black lines for a UPC triggered event. 
Other events recorded in the TPC are shown with different colors.}
\end{minipage}\hspace{4pc}
\begin{minipage}{18pc}
\includegraphics[width=13pc, angle=90]{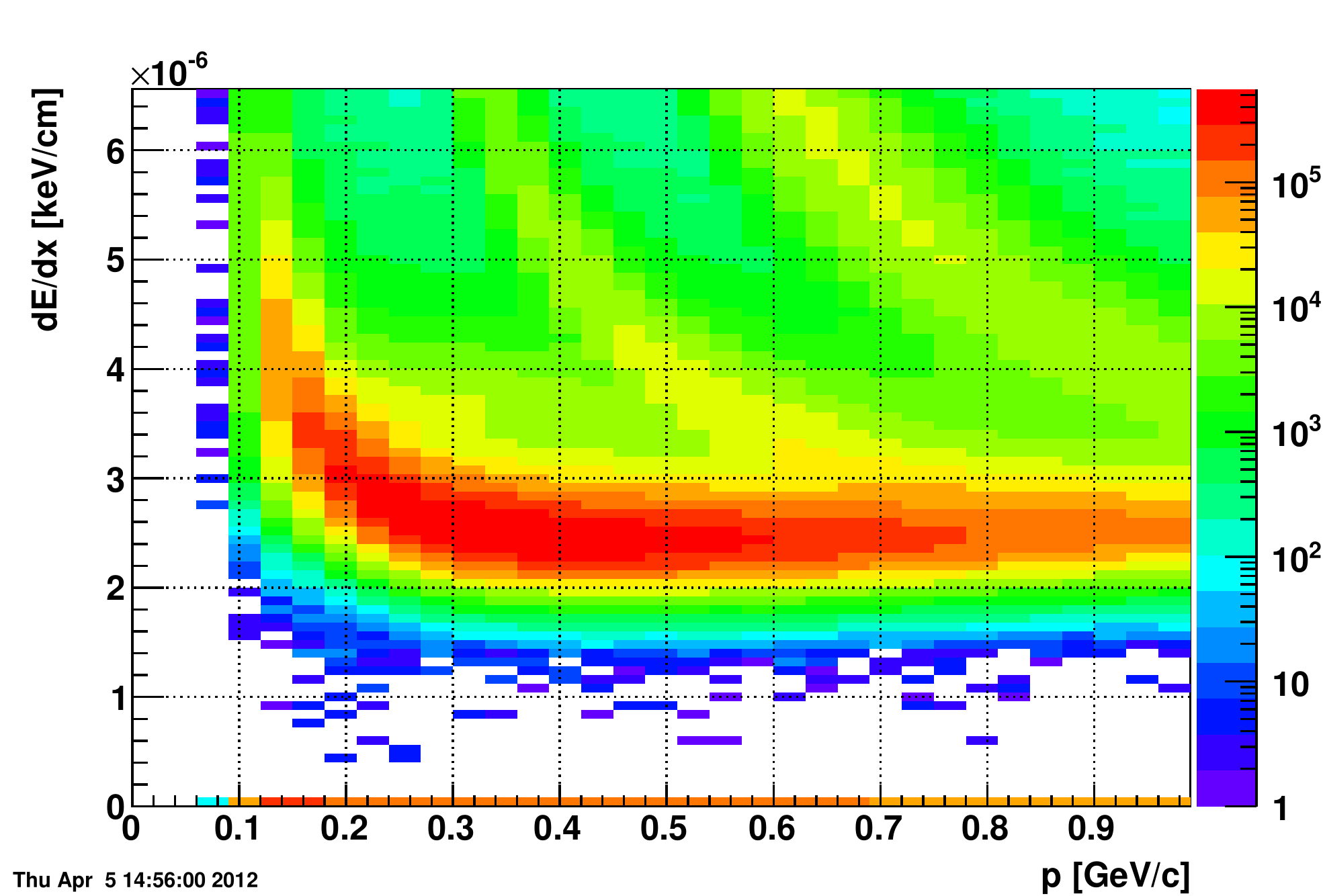}
\caption{\label{dEdx}The average ioniation energy loss (in keV/cm) recorded for tracks detected in the TPC is shown as function of
their measured momentum (in GeV/c). The pion band is dominant in this figure but both kaon and proton bands are also visible.}
\end{minipage}\hspace{8pc}%
\end{figure}

During the third and final pass on the data, events are first selected to have no more that one neutron in each one of the ZDCs, this 
selects the 1n1n exitation of the ions. Figure \ref{zdcCuts} shows the pedestal subtracted signal from the East ZDC, The cut for a single
neutron is placed at ADC channel 300.
Tracks that are part 
of the triggered events are identified through their vertex and
the one associated to the TOF hits present in the data. Pairs that do not have that connection to the TOF hits that triggered the 
events are rejected. Figure \ref{manyEvents} shows one event in the YZ view of the TPC, tracks originating from the same vertex are
presented as lines of the same color. The tracks shown with black lines are the ones connected to at least one TOF hit.

\begin{figure}[h]
\begin{minipage}{16pc}
\includegraphics[width=16pc]{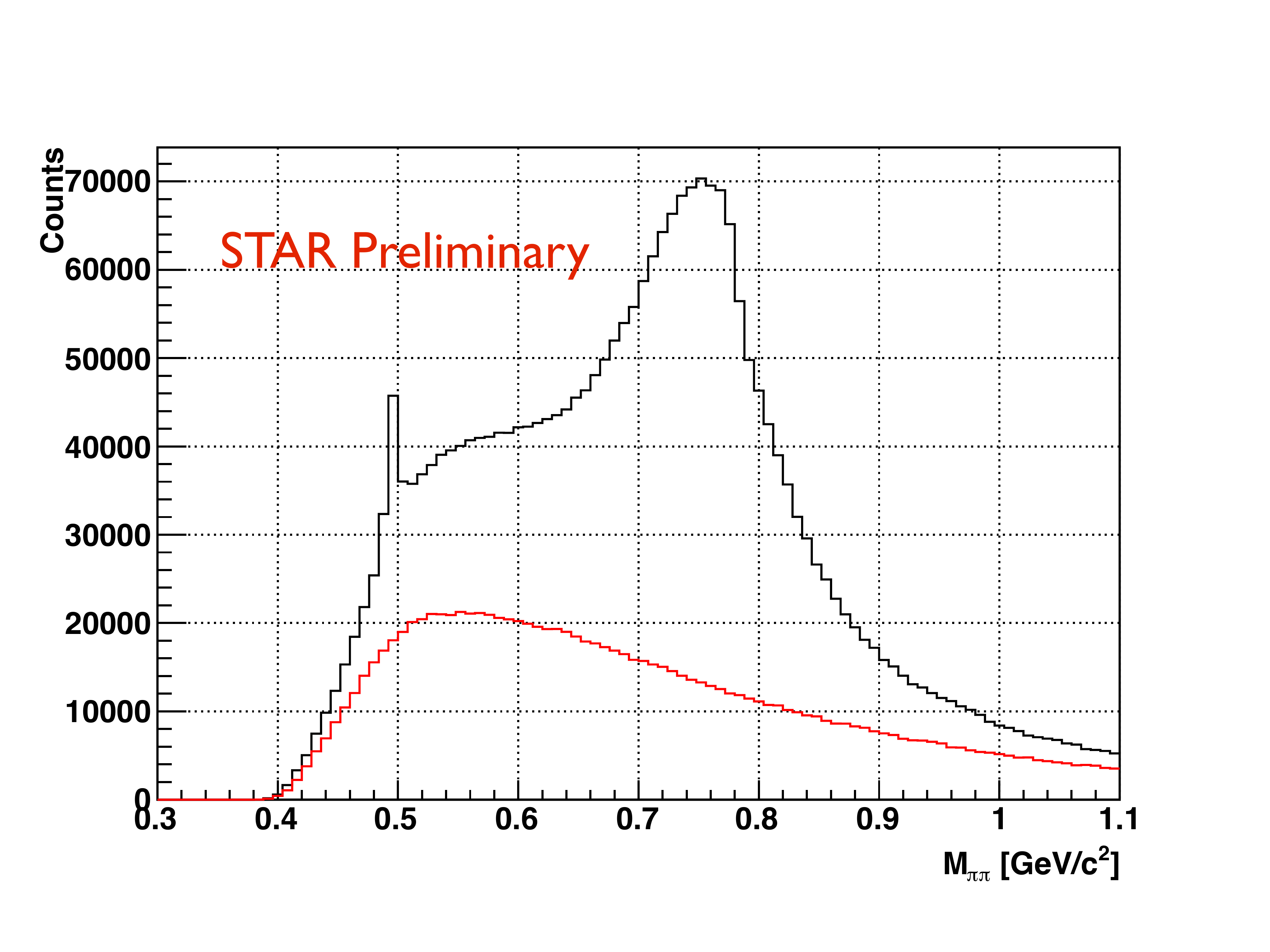}
\caption{\label{invMass}Invariant mass distribution for pion pairs of opposite charge shown with a black
histograms, together with the similar distribution filled with pion pairs of equal charge.}
\end{minipage}\hspace{4pc}%
\begin{minipage}{18pc}
\includegraphics[width=16pc]{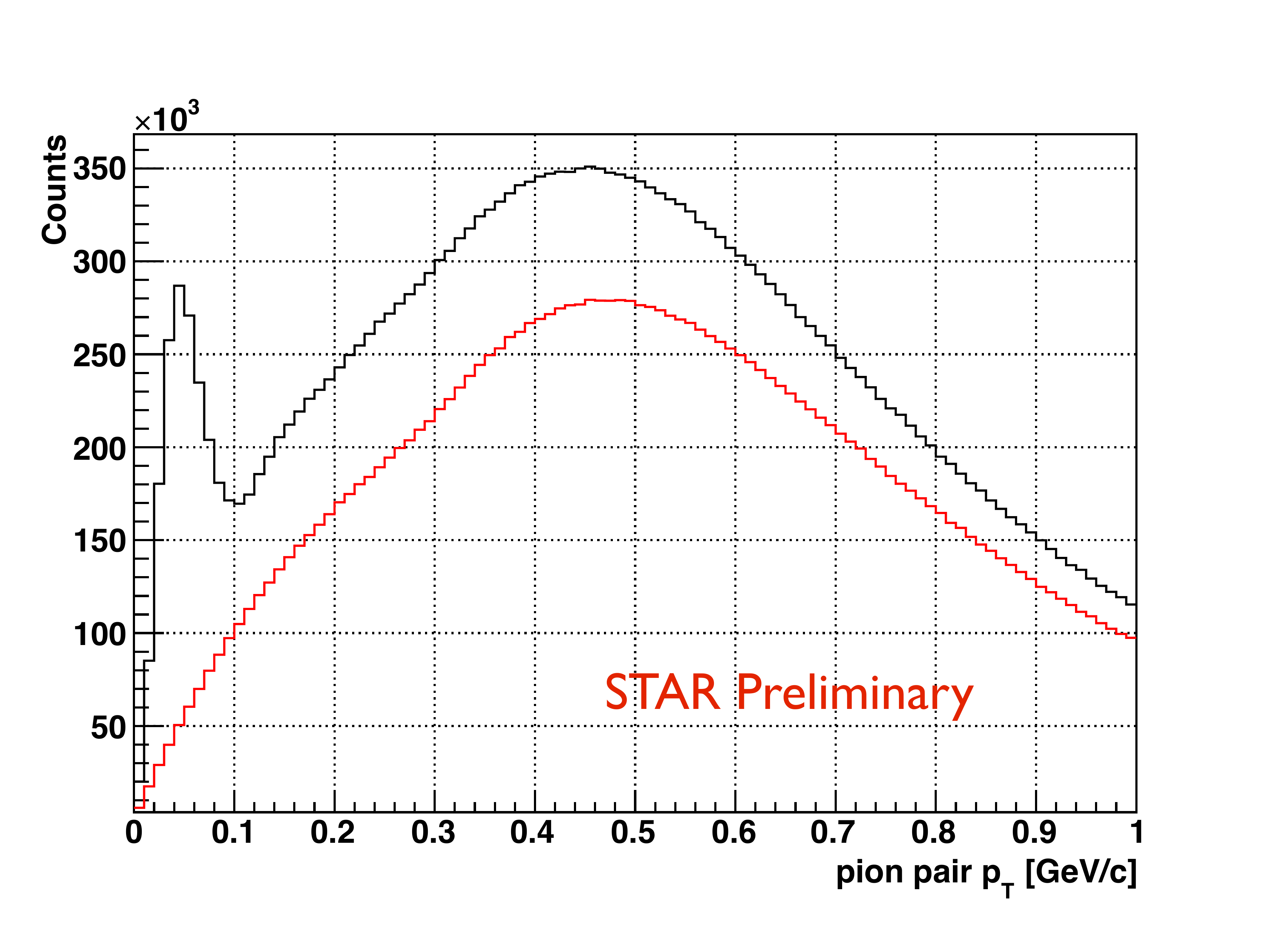}
\caption{\label{pairPt}Transverse momentum distribution of pairs of opposite charge are shown with a black histogram. The red histogram 
shows similar distribution but this time for pion pairs of equal charge.}
\end{minipage} 
\end{figure}

The identity of each track in the selected pairs is obtained from the amount of ionization energy deposited along their trajectory inside
the TPC volume. Tracks that have less than 14 samples (hits) are eliminated from the set because they are found to be produced by low 
momentum particles spiraling along the solenoidal magnetic field. The resolution of the $dE/dx$ measurement in the TPC is very good
and allows a clean identification of pions up to $\sim 1~ GeV/c$ covering the range of momenta for pions originating from $\rho$ mesons 
produced practically at rest. Figure \ref{dEdx} shows the correlation between  $dE/dx$ and the momentum of the particles. Pions
are clearly visible as the mostly horizontal band of ``minimum ionization''. With this identification it is now possible to select events
that contain pion pairs. Pairs with different charge will then be considered as candidates decays from a $\rho^{0}$ meson. Pairs of tracks 
with equal charge $++$ or $ --$ are also used to build our best estimate so far of the background for the production of $\rho^{0}$ mesons in
UPC events. The invariant mass of both types of pairs is constructed and is shown in Fig. \ref{invMass} where the black histogram
shows the mass distribution for pairs of opposite charge, and the red one shows the mass distribution from equal charge pairs, which from
now on will be considered as the background to this $\rho^{0}$ meson study. The wide mass distribution centered at 770 MeV is dominant 
in the 
black histogram, as well as a very narrow peak at the kaon mass, most likely produced by the decay of $K_{S}^{0}$ into two pions.

Figure \ref{pairPt} shows the distribution of the transverse momentum of the pion pair for opposite charge pions in the black histograms
and the same charge pion pairs shown in the red histogram. The distribution shows a prominent peak below 100 MeV/c and a much broader 
distribution at higher value of $p_{T}$. The low transverse momentum peak is related to $\rho$ mesons scattering coherently off the Au 
ion. In other words, the uncertainty principle connects the narrow low \pT~ peak with a broad spatial structure; the Au nucleus.    
An earlier STAR publication \cite{prl89} dealing with UPC $\rho^{0}$ meson coherent production imposed a cut on the $p_{T}$ of the pion 
pair at 150 MeV/c
to select coherent scattering, in this analysis no such cut is applied and a subtraction of the incoherent component of the $\rho$ 
scattering is used to enhance the coherent events. The assumption made is that the coherent component has a rapidly falling tail that 
extends to higher values of $p_{T}$ and is completely hidden in Fig. \ref{pairPt} by the less steeply falling incoherent component
 as well as by the increasing contribution of background contamination. The best estimate of the pion background in our data sample is the
corresponding shape and magnitude of the equal sign pairs in every studied distribution, as done in the STAR publication \cite{prc77}.

The last condition applied to the pairs on this third pass requires that the vertex of the selected event is formed with only  two tracks. 
This
condition, together with the veto imposed on the BBC detectors should then make our sample of $\rho^{0}$ mesons as exclusively produced
in UPC events.

\begin{figure}[h]
\begin{minipage}{17pc}
\includegraphics[width=18pc]{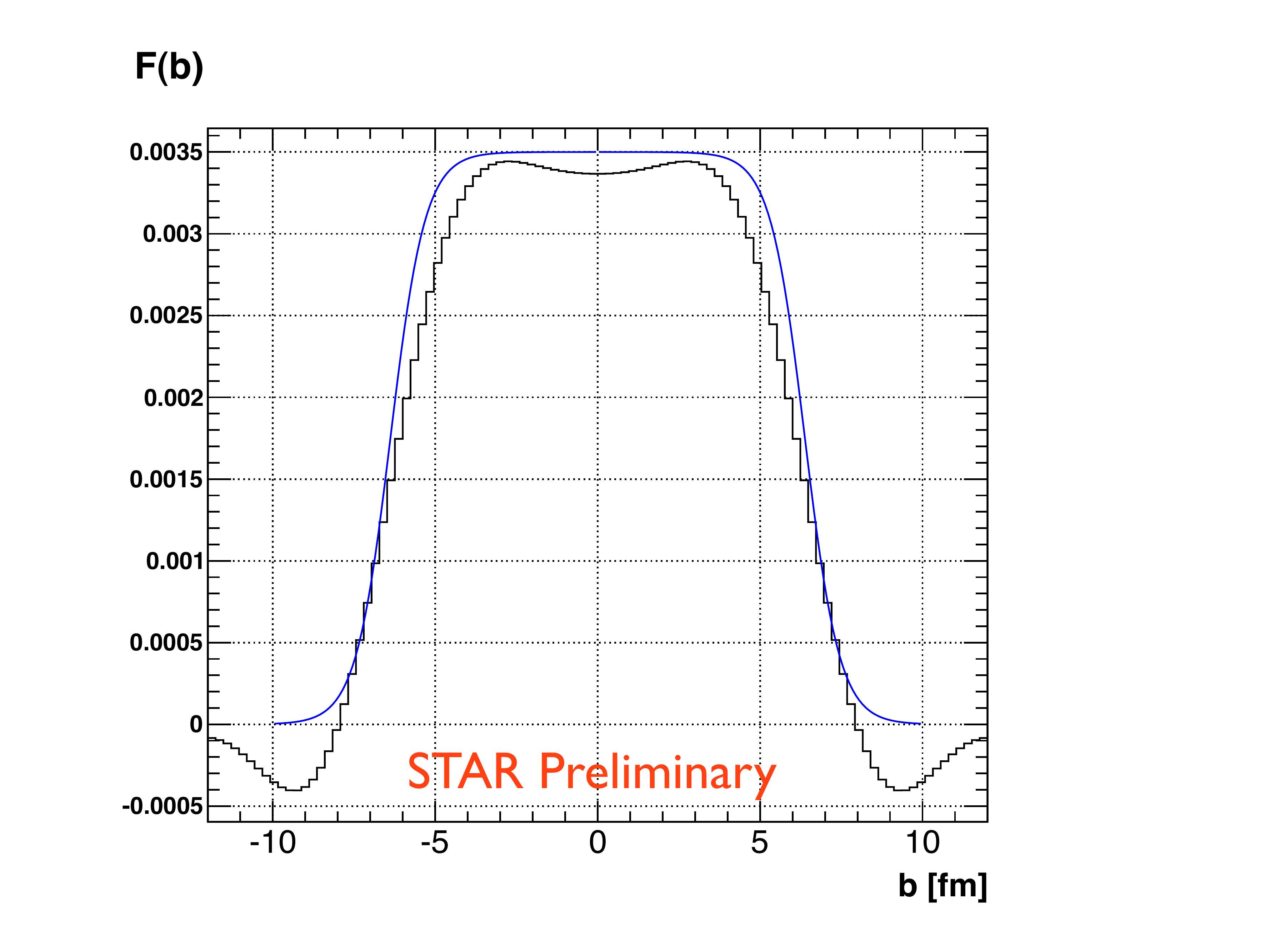}
\caption{\label{fourier}One dimensional Fourier transformation of the measured diffraction pattern shown a a black histogram. The Wood Saxon 
functional form with R=6.38 fm and $a_{0}=0.535~ fm$ is shown with a blue curve. }
\end{minipage}\hspace{3pc}%
\begin{minipage}{18pc}
\includegraphics[width=19pc]{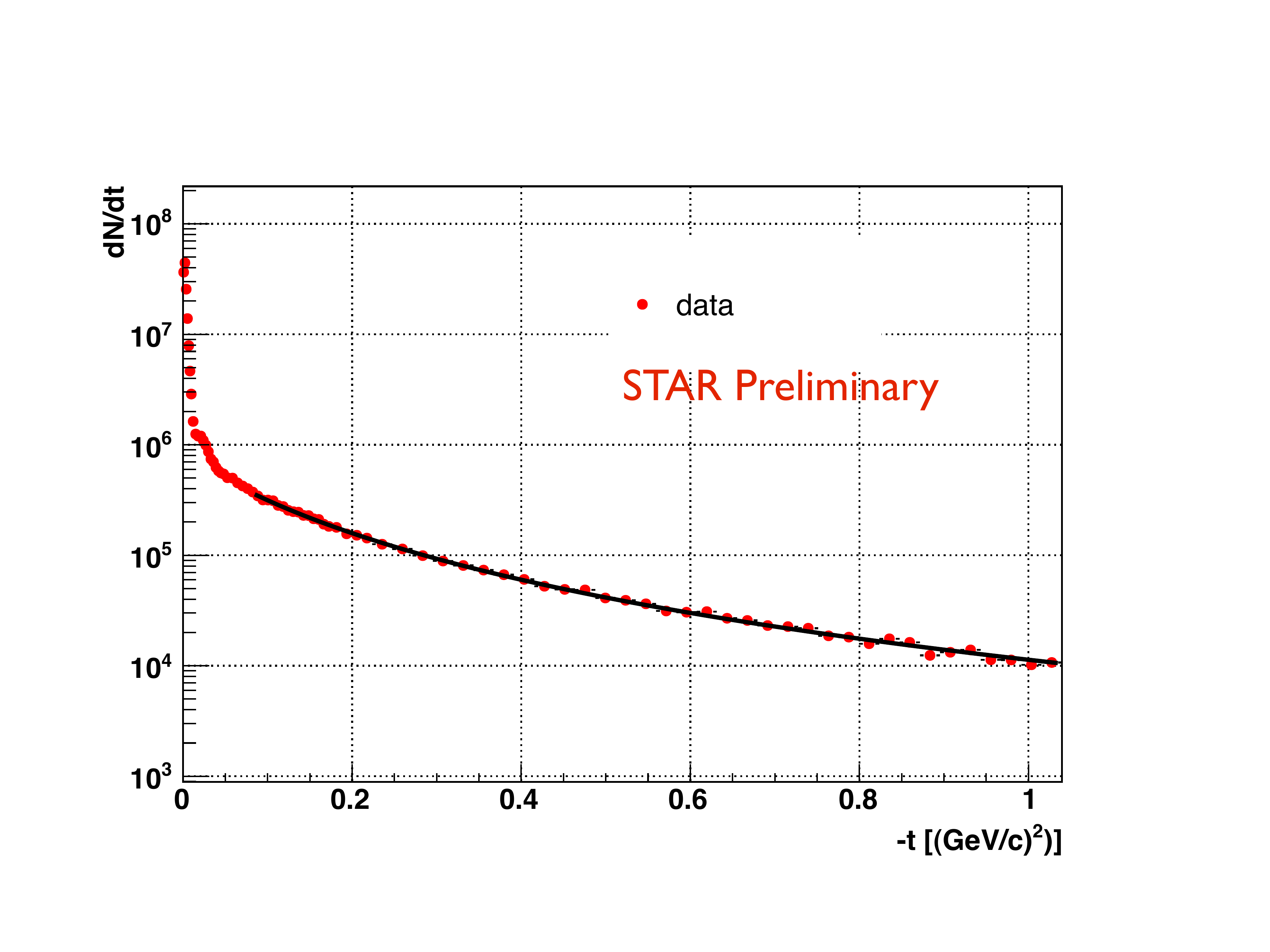}
\caption{\label{bkgSub}Background subtracted distribution of $\rho$ mesons as function of -t shown with red markers. A fit to a power law
function is shown with a black curve. No acceptance correction is applied.}
\end{minipage} 
\end{figure}

The $\rho$ meson is finally defined with cuts applied to the invariant mass distribution of the pion pairs that satified the
conditions listed above. The lowest value cut on mass is set at 500 MeV and the top value of the same cut at 1 GeV.

\section{The diffraction pattern}

We make the important assumption that the t of the $\rho^{0}$ meson completely balances the one of the target Au ion and
what we are producing is the t distribution of the recoil ion. The appropriateness of such assumption will be discussed again
when we compare the measurements to model calculations. 

The t Mandelstam variable for the $\rho$ defined as $t=(P_{1} - P_{3})^2$ where the four-vectors $P_{1}$ and $P_{3}$ are related to
the $\rho$ meson before and after the elastic interaction with the Au ion. In the lab. frame this quantity is written as 
$t = t_{min} - p_{T}^{2}$ where $t_{min}= -M_{\pi^{+}\pi^{-}}^{2}/(2\gamma_{L} e^{|2y|})$ with $\gamma_{L}$ being the Lorentz boost of the 
ion beam (equal to 108 at
top RHIC ion energy) and y the rapidity of the $\rho$ mesons. The value of $t_{min}$ is small and was included 
for completeness. 

\begin{figure}[h]
\begin{center}
\includegraphics[width=28pc]{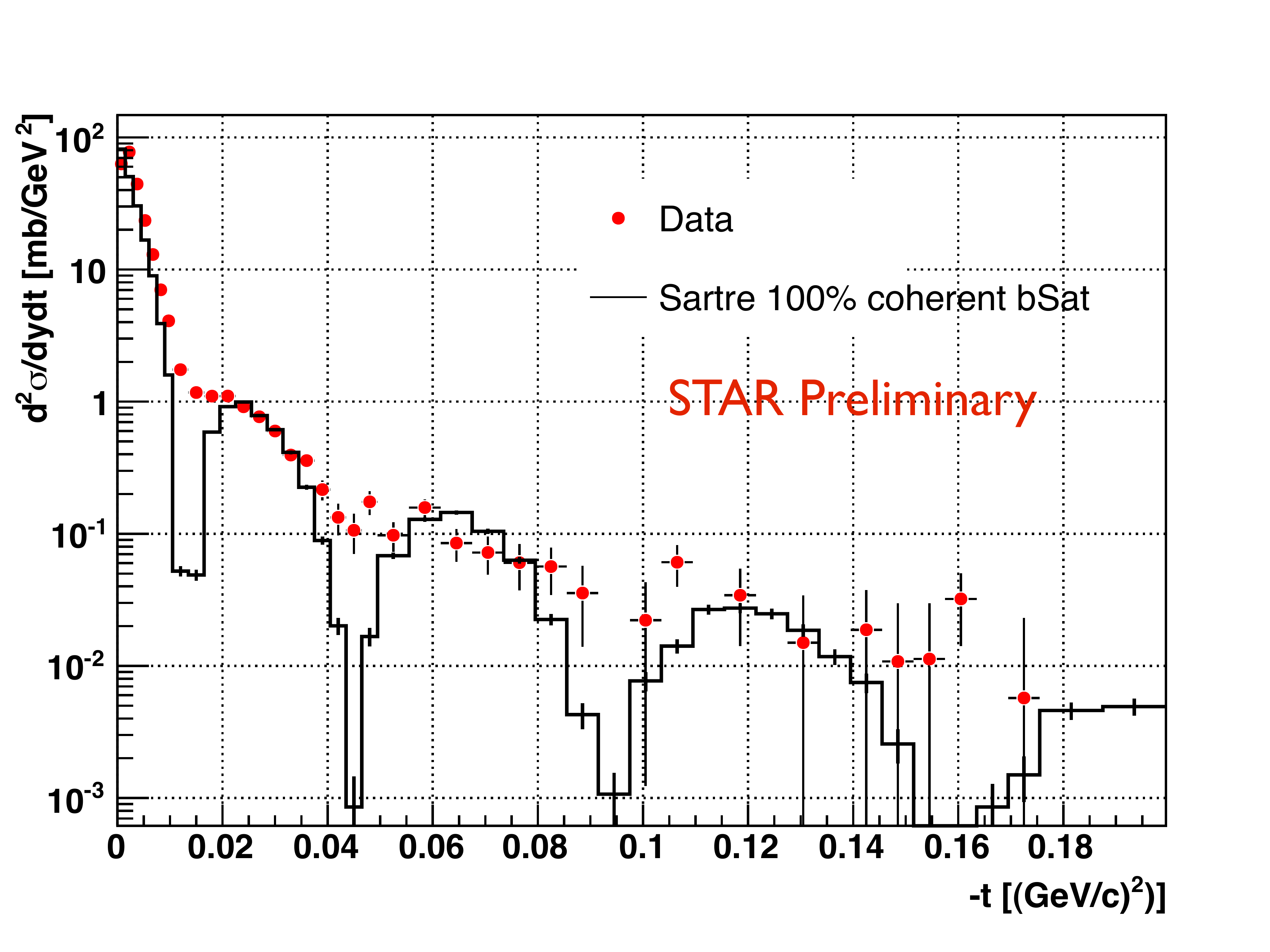}
\caption{\label{finalResult}The diffraction pattern extracted from elastic scattering from the Au nuclei 
shown with red markers. The black histogram is the result of a calculation performed with Sartre.}
\end{center}
\end{figure}

Two histograms with variable bin size are prepared to be filled with the value of t for the exclusively produced $\rho$ mesons detected
in events with GDR excitations in both Au ions recording a single neutron in each ZDC. The bin width ranges from 0.0015 $(GeV/c)^{2}$ 
at $t\sim 0$ to 0.024 $(GeV/c)^{2}$ at 
$t=0.3 (GeV/c)^{2}$. The bin size was chosen to be always wider than the corresponding t resolution ($\Delta t/t = 0.01 + 0.008\sqrt{t}$).
One histogram ($dN_{\rho}^ {candidates}/dt$) is used to record the -t distribution of opposite sign pions pairs. The other histogram
($dN_{background}/dt$) is filled with the value of -t for pion pairs of equal charge. As mentioned above, this second histogram is 
considered to carry all the information about the backgroung to $\rho$ photo-production in UPC events. Studies to characterize the 
validity of this choice are still under way.
At every bin, the contents of both histograms are divided by the corresponding bin width to produce the actual distributions. 
 The $dN_{background}/dt$ distribution is then subtracted bin by bin from $dN_{\rho}^ {candidates}/dt$ and 
the statistical errors are recalculated. The resulting distribution ($dN_{\rho}^ {signal}/dt$) is shown with red dots 
in Fig. \ref{bkgSub} where 
the first diffractive peak is
clearly visible at $-t \sim 0$, a clear hint of the second diffractive peak is also present at this point in the analysis. A long slowly
falling tail is present at higher values of -t. At this point the assumption is made that the coherent component of the $\rho$ meson
scattering off the Au nuclei is negligeable and the totality of this tail is populated by incoherent interactions with individual
nucleon. A fit to a ``power law'' shape $(A/(1+|t|/p_{0})^{n})$  is performed for $-t > 0.2~(GeV/c)^{2}$ This functional form reproduces
well the incoherent tail ($\chi /NDF = 0.97~ A=7.609\times10^{4}, p_{0} = 0.23\pm0.1~ {\rm and}~ n=2.8\pm0.4$). The separation 
between coherent and
incoherent contributions to the elastic scattering of the $\rho$ meson on the Au nuclei is done statistically by subtracting the value
of the fitted function at each bin center of the background subtracted distribution ($dN_{\rho}^ {signal}/dt$). The resulting
distribution is shown in Fig. \ref{finalResult} where all normalization and other corrections were applied.

The $\rho$ meson reconstruction efficiency and the magnitude of the TPC acceptance are extracted making use of of STAR embedding 
procedures; pion pairs from the decay of one million $\rho^{0}$ mesons generated in UPC events with the StarLight \cite{baltzStarLight} 
event generator are simulated and their trajectories and energy deposited along them are produced with a complete and up to date 
GEANT simulation of the STAR detector setup as it was at the time of  RHIC run 10. The hits recorded in this simulation are then
merged with information recorded during actual data taking in run 10 with the most unbiased trigger (zerobias). Embedding into
these events produces the best description of the backgroung in UPC events; a mix of pile-up, beam gas events as well as low 
multiplicity hadronic interactions. The complete STAR reconstruction chain is then used to process these data and the result is
presented in several formats. To extract a first version of the reconstruction efficiency and the acceptance, a summary  output 
was used where the information about the input generated particles as well as the parameters of the reconstructed tracks that match
to a generated particle. The matching between reconstructed tracks and generated particles is based on the number of common hits.
Ratios between 2D histograms counting matched and generated particles were calculated for different combinations of the $\rho$ meson
 variables:
transverse momentum and rapidity, azimuthal angle and rapidity and -t and rapidity. The value of efficiencies$\times$acceptance 
as function of -t was extracted. For the study reported here the rapidity dependance is averaged and the resulting value of the 
correction is equal to 44\% and is found to be flat in -t.


The integrated luminosity for the runs included in this analysis is equal to $679 (\mu b)^{-1}$ and is based on one of the STAR 
minimum bias triggers ($minbias\_monitor$)
which is estimated to cover up to 6 barns of the total Au+Au cross section.


\subsection{Comparison to model calculations}

\begin{figure}[h]
\begin{minipage}{18pc}
\includegraphics[width=19pc]{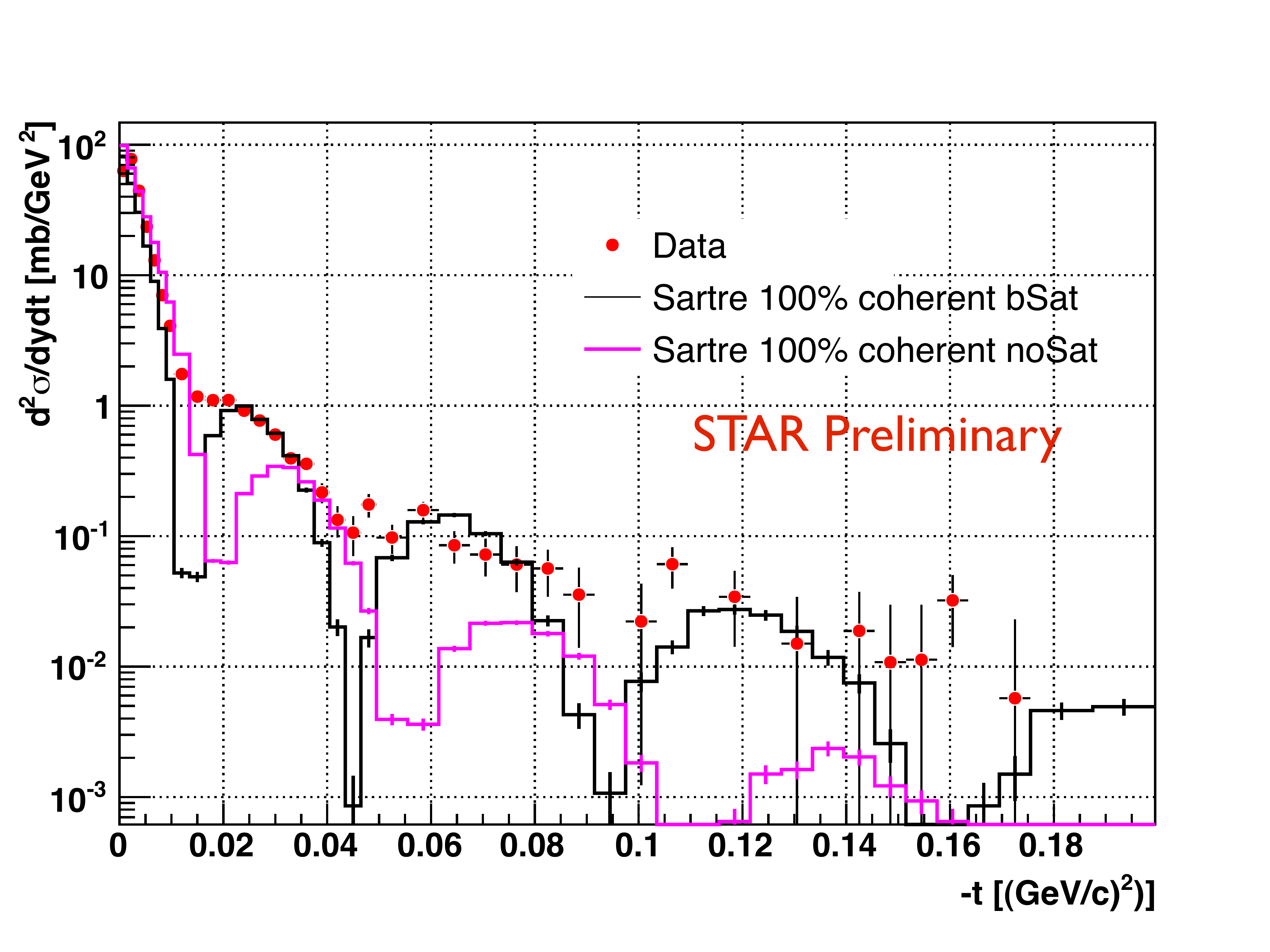}
\caption{\label{nonSat}Diffraction patter shown with two Sartre calculations, one with the presence of 
saturation in the gluon distribution of the ion target shown as a black histogram, and the other were
such non linear effect have been turned off, is shown with a magenta histogram.}
\end{minipage}\hspace{2pc}%
\begin{minipage}{18pc}
\includegraphics[width=19pc]{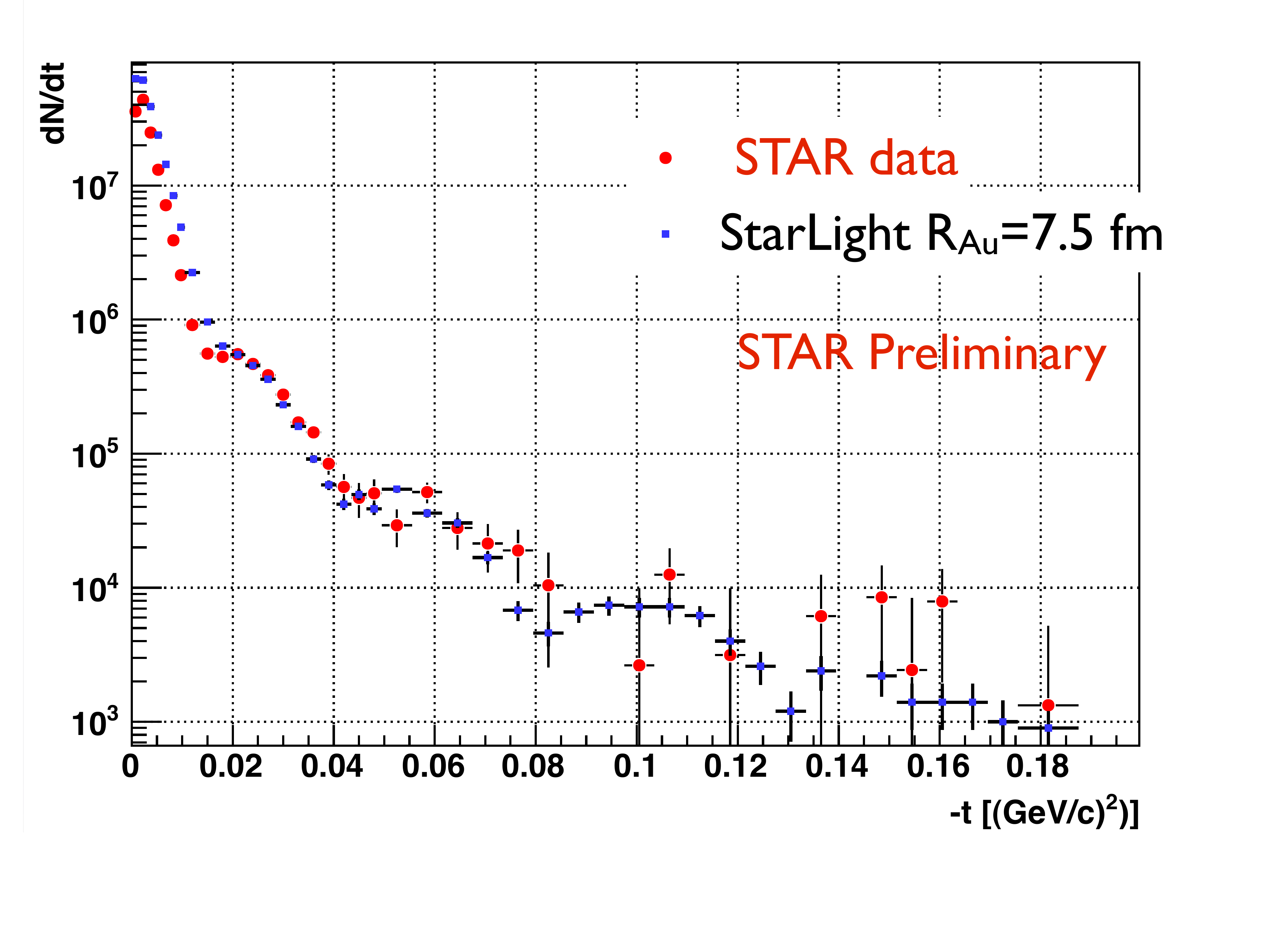}
\caption{\label{starlightComp}The unnormalized distribution for the -t value of $\rho$ mesons generated with Starlight is shown with blue markers
to compare them with the same distribution extracted from data. The comparisons highlights the good match of peaks and valleys in both cases.
}
\end{minipage} 
\end{figure}

The histogram shown in Fig. \ref{finalResult} was produced with the Sartre event generator \cite{sartre} which in turn is based on the
impact parameter dependent saturation model bSat \cite{Kowalski:2003hm}. Ten million $\rho$ mesons were generated with the model 
set to mimic
the Au+Au Ultraperipheral collisions at RHIC. The geometry of the target Au ion is simulated in a Glauber based Monte Carlo with 
individual nucleons distributed according to a Wood Saxon distribution with R=6.38 fm and $a_{0}=0.535~ fm$. The cross section of the
quark dipole is connected to the gluon distribution in the target nuclei and in particular, the black histogram in Fig. \ref{finalResult}
includes the effects of saturation.
 The agreement
between the data and the simulation is remarkable. The slopes of both first peaks are close to each other, the location and
magnitude of the other diffraction peaks formed in the measurement are also well reproduced by the calculation.
To first order, this agreement is strong indication that the methods used to isolate the coherent component of the $\rho$ 
elastic scattering in the present analysis
were correct, even though a quantification of such agreement remains to be done.      
A change in the di-quark cross section in the Sartre event generator which this time does not include the effects of saturation produces
strong changes in the position of the deeps in the calculated diffraction pattern as is shown with the magenta histogram in Fig. \ref{nonSat}

The Fourier transform of the measured diffraction pattern has been calculated making use of the azimuthal symmetry of the system
: $f(b) = \int_0^{2\sqrt{t_{max}}} \sqrt{F(x^{2})} J_{0} (xb/\hbar c) \frac{ x}{2\pi} dx$
The result of the transformation is shown as a histogram in Fig. \ref{fourier}, together with the outline
of the Wood-Saxon distribution drawn with a blue curve. No attempt has been made yet at assigning a systematic
uncertainty to this calculation. The high b sides of the transform are related to $-t \sim 0$ in the diffraction
pattern and the fact that they take negative values may be an artifact of bin size or the presence of 
destructive interference already measured in the same system \cite{prl102}. The region close to b=0 is not well defined 
because the measurement has poor statistics at high -t. The falling edges of the distribution will turn out to be the 
best suited to extract some physics out of this measurement.
Figure \ref{starlightComp} shows the comparison between the diffraction pattern extracted from data (red markers) and the one caculated
with the Starlight event generator (blue markers). This comparison shows a remarkable agreement on the depths of the diffraction deeps 
which within the
Starlight calculation are the result of non zero transverse momentum component in the Weizs\"{a}cker-Williams photon exchanged between 
the ions \cite{starLightDeeps}. A similar calculation performed for UPC events at 130 GeV produced shallow diffraction deeps 
\cite{Strickman}.  Such facts make our original assumption of purelly longitudinal photons too simplistic but it does not
preclude the possible unfolding of the photons transverse momentum contribution from our present results.

\section{Summary}


The STAR UPC program has measured coherent elastic scattering off Au nuclei with beams of exclusively 
produced
$\rho^{0}$ mesons in Ultra Peripheral Collisions of Au ions at \snn = 200 GeV. The diffraction pattern extracted using the \pT~ of the
scattered meson may not be an exact match to the one measured with the recoiling target ion, but such goal appears attainable with the
help of a QED calculation to unfold the transverse contribution from the Weizs\"{a}cker-Williams photons.  This measurement opens up a 
possible program to study the gluon form factor of nuclei, albeit with a restricted
reach in $Q^{2}$.

\section*{References}
\bibliography{RhoDiffractionWWND2012}

\providecommand{\newblock}{}
\begin{thebibliography}{10}
\expandafter\ifx\csname url\endcsname\relax
  \def\url#1{{\tt #1}}\fi
\expandafter\ifx\csname urlprefix\endcsname\relax\def\urlprefix{URL }\fi
\providecommand{\eprint}[2][]{\url{#2}}

\bibitem{Bertulani}
Bertulani C {\em et~al.\/} 2005 {\em Ann. Rev. Nucl. Part. Phys.\/} {\bf 55}
  271

\bibitem{busza}
Bulos F {\em et~al.\/} 1969 {\em Phys. Rev. Lett.\/} {\bf 22} 490

\bibitem{samTing}
Alvensleben H {\em et~al.\/} 1970 {\em Phys. Rev. Lett.\/} {\bf 24} 786

\bibitem{TPCnim}
Anderson M {\em et~al.\/} 2003 {\em Nuclear Instrum. Methods\/} {\bf A499} 659

\bibitem{Adler:2001fq}
Adler C, Strobele H, Denisov A, Garcia E, Murray M {\em et~al.\/} 2001 {\em
  Nucl.Instrum.Meth.\/} {\bf A461} 337--340

\bibitem{baltzStarLight}
Baltz A~Klein S and Nystrand J 2002 {\em Phys. Rev. Lett.\/} {\bf 89} 012301

\bibitem{prl89}
Adler C {\em et~al.\/} 2002 {\em Phys. Rev. Lett.\/} {\bf 89} 272302

\bibitem{prc77}
Abelev B~I {\em et~al.\/} (STAR) 2008 {\em Phys. Rev.\/} {\bf C77} 034910
  (\textit{Preprint} \eprint{0712.3320})

\bibitem{sartre}
Toll T and Ulrich T 2011 vol INT-PUB-11-034 p 385 (\textit{Preprint}
  \eprint{nucl-th/1108.1713})

\bibitem{Kowalski:2003hm}
Kowalski H and Teaney D 2003 {\em Phys.Rev.\/} {\bf D68} 114005
  (\textit{Preprint} \eprint{hep-ph/0304189})

\bibitem{prl102}
Abelev B {\em et~al.\/} 2009 {\em Phys. Rev. Lett.\/} {\bf 102} 112301

\bibitem{starLightDeeps}
Klein S and Nystrand J 2000 {\em Phys. Rev. Lett.\/} {\bf 84} 2330

\bibitem{Strickman}
Frankfurt L, Strikman M and Zhalov M 2003 {\em Phys. Rev.\/} {\bf C67} 034901

\end{thebibliography}
\bibliographystyle{iopart-num}

\end{document}